\documentclass[12pt]{article}
\usepackage{amsmath}
\usepackage{amssymb}

\newcommand{\cl}{\mathcal{L}}
\newcommand{\ce}{\mathcal{E}}
\newcommand{\ha}{H^{(\alpha)}}
\newcommand{\ea}{E^{(\alpha)}}
\newcommand{\psia}{\psi^{(\alpha)}}
\newcommand{\veff}{V^{(\rm eff)}}
\newcommand{\cna}{{\cal N}^{(\alpha)}}
\newcommand{\ka}{K^{(\alpha)}}
\newcommand{\Aa}{A^{(\alpha)}}
\newcommand{\ca}{C^{(\alpha)}}
\newcommand{\phia}{\phi^{(\alpha)}}
\newcommand{\epsilona}{\epsilon^{(\alpha)}}
\newcommand{\ma}{M^{(\alpha)}}
\newcommand{\chia}{\chi^{(\alpha)}}
\newcommand{\cea}{\mathcal{E}^{(\alpha)}}
\newcommand{\na}{N^{(\alpha)}}

\title{
Oscillator-Morse-Coulomb mappings and algebras for constant or position-dependent mass}
\author{C Quesne\thanks{Electronic mail: cquesne@ulb.ac.be}\\ 
{\small\sl Physique Nucl\'eaire Th\'eorique et Physique Math\'ematique,  Universit\'e Libre de Bruxelles,} \\ 
{\small\sl Campus de la Plaine CP229, Boulevard~du Triomphe, B-1050 Brussels, Belgium}}
\date{ }
\begin{document}
\baselineskip=22pt plus 1pt minus 1pt
\maketitle
\begin{abstract}
The bound-state solutions and the su(1,1) description of the $d$-dimensional radial harmonic oscillator, the Morse and the $D$-dimensional radial Coulomb Schr\"odinger equations are reviewed in a unified way using the point canonical transformation method. It is established that the spectrum generating su(1,1) algebra for the first problem is converted into a potential algebra for the remaining two. This analysis is then extended to Schr\"odinger equations containing some position-dependent mass. The deformed su(1,1) construction recently achieved for a $d$-dimensional radial harmonic oscillator is easily extended to the Morse and Coulomb potentials. In the last two cases, the equivalence between the resulting deformed su(1,1) potential algebra approach and a previous deformed shape invariance one generalizes to a position-dependent mass background a well-known relationship in the context of constant mass.  
\end{abstract}

\noindent
Running title: Oscillator-Morse-Coulomb mappings and algebras


%
%
\newpage
\section{INTRODUCTION}

Since the pioneering work of Pauli on the hydrogen atom energy spectrum \cite{pauli}, there has been a continuing interest in applying Lie algebraic methods to deriving the bound-state eigenvalues and eigenfunctions of quantum systems. In this respect, the su(1,1) algebra has attracted considerable attention \cite{wybourne}, partly due to its smallness and its well-known unitary irreducible representations (unirreps) \cite{barut65}.\par
%
%
In most applications \cite{footnote1, delsol}, su(1,1) appears either as a spectrum generating algebra \cite{barut88} or as a potential algebra \cite{alhassid}. In the former case, the Hamiltonian is directly  related to the su(1,1) weight generator, while the raising and lowering generators are ladder operators connecting eigenstates of the Hamiltonian with different eigenvalues. In the latter case, on the contrary, the raising and lowering generators act as shift operators between pairs of eigenstates belonging to a hierarchy of Hamiltonians of a given family, but characterized by the same energy, which is related to the Casimir operator eigenvalue. This approach has been shown \cite{gango} to be equivalent to that based on supersymmetric quantum mechanics (SUSYQM) for translational shape-invariant potentials (see also Ref.\ \cite{bagchi95}).\par
%
%
In two recent papers \cite{cordero, martinez}, the occurrence of the su(1,1) algebra has been re-examined in three non-relativistic quantum mechanical problems, namely the radial harmonic oscillator, the Morse and the radial Coulomb ones. Although these three topics have been the subject of many works (see Refs.\ \cite{cordero} and \cite{martinez} for lists of references), there still remain some points at issue in their analysis. Among them, let us mention a unified treatment of these potentials and of their respective su(1,1) algebras, an unambiguous identification of the role played by the latter ({\sl i.e.}, spectrum generating {\sl vs.} potential algebra), and an extension of the picture to some position-dependent mass (PDM) cases. It is the purpose of the present paper to provide replies to these questions.\par
%
%
The three potentials in hand belong to the so-called Natanzon confluent potential class \cite{natanzon}. The corresponding Schr\"odinger equations can therefore be related through point canonical transformations (PCT) \cite{bhatta}. These have been extensively used in the SUSYQM framework \cite{kostelecky, haymaker}, but also in other contexts \cite{moshinsky72}. Such a method will allow us to directly derive the su(1,1) algebras for the Morse and Coulomb potentials from the well-known spectrum generating algebra obtained in the oscillator case \cite{wybourne}.\par
%
%
On the other hand, much interest has been recently devoted to finding exact solutions to Schr\"odinger equations involving known potentials, {\sl e.g.}, oscillator, Morse or Coulomb, in a PDM context \cite{dekar, milanovic, plastino, dutra, roy, alhaidari, gonul, koc, bagchi04, cq04, yu, chen, bagchi05, dong, jiang, cq06, mustafa, tanaka, ganguly, carinena, schulze, cq07}. There indeed exists a wide variety of physical problems in which an effective mass depending on the position is of utmost relevance \cite{weisbuch}. PDM also holds out to deformations in the quantum canonical commutation relations or curvature of the underlying space \cite{cq04, carinena}. Furthermore, it has recently been observed \cite{bagchi06} that there exists a whole class of Hermitian PDM Hamiltonians which, to lowest order of perturbation theory, have a correspondence with pseudo-Hermitian Hamiltonians, whose study is a topic of current considerable interest \cite{bender}.\par
%
%
It may be worth mentioning that in the classical limit, the quantum Hamiltonians contained in PDM Schr\"odinger equations go over to classical Hamiltonians with a similar PDM \cite{daboul}. On the other hand, the quantization problem for such classical Hamiltonians has been the subject of many works, due to the momentum and mass-operator noncommutativity and the resulting ordering ambiguity in the kinetic energy term (for more details we refer the reader to the references given in \cite{bagchi04, cq04, cq06}).\par
%
%
Some examples of PDM Schr\"odinger equations have been presented in Ref.\ \cite{bagchi05}, where bound-state spectra have been obtained for a variety of potential and PDM pairs by appropriately deforming the SUSYQM shape-invariance condition under parameter translation. In a recent work \cite{cq07}, we have considered the PDM found there for the $d$-dimensional radial harmonic oscillator potential  and we have constructed a spectrum generating algebra for the corresponding Schr\"odinger equation. Using a nontrivial approach based upon quadratic algebras, we have succeeded in obtaining a deformed su(1,1) algebra, which goes over to the standard one in the limit where the mass becomes constant. As far as we know, this is the first case where a spectrum generating algebra has been built in a PDM background.\par
%
%
It is therefore of utmost relevance to enquire into other examples. The PCT approach, considered for constant mass, will actually provide us with a key to solving similar problems for the Morse and Coulomb potentials with some appropriately chosen PDM without having to resort to the complicated analysis carried out for the harmonic oscillator.\par
%
%
This paper is organized as follows. In Sec.~II, the radial oscillator, Morse and radial Coulomb problems are reviewed for a constant mass in the PCT framework. This approach is then extended to a PDM context in Sec.~III. Finally, Sec.~IV contains the conclusion.\par
%
%
\section{MAPPINGS AND ALGEBRAS FOR CONSTANT MASS}

\subsection{Oscillator case}

Let us start from the radial Schr\"odinger equation for the $d$-dimensional harmonic oscillator ($d \ge 2$), which can be written as
\begin{gather}
  H_{\rm HO} \psi(r) = E \psi(r),  \label{eq:HO-SE} \\
  H_{\rm HO} = - \frac{d^2}{dr^2} + \frac{L(L+1)}{r^2} + \frac{1}{4} \omega^2 r^2,  \label{eq:HO-H}
\end{gather}
in units wherein $\hbar = 1$ and the mass $m_0 = 1/2$. Here $r$ runs on the half-line $0 < r < \infty$ and $L$ is defined by $L = l + (d-3)/2$ in terms of the angular momentum quantum number $l$. As we have eliminated the first-order derivative in (\ref{eq:HO-H}), the radial wavefunction is actually $r^{-(d-1)/2} \psi(r)$, so that the normalization condition for $\psi(r)$ reads
\begin{equation}
  \int_0^\infty |\psi(r)|^2 dr = 1.  \label{eq:HO-sp}
\end{equation}
\par
%
%
Equation (\ref{eq:HO-SE}) has an infinite number of bound-state solutions~\cite{moshinsky72, moshinsky96}
\begin{equation}
  \psi_{n,L}(r) = {\cal N}_{n,L} r^{L+1} e^{-\frac{1}{4} \omega r^2} L_n^{(L+\frac{1}{2})}(\tfrac{1}{2} \omega
   r^2), \qquad n = 0, 1, 2, \ldots,  \label{eq:HO-psi}
\end{equation}
corresponding to the energy eigenvalues
\begin{equation*}
  E_{n,L} = \omega (2n + L + \tfrac{3}{2}).  
\end{equation*}
In (\ref{eq:HO-psi}), $L_n^{(\alpha)}(y)$ denotes a Laguerre polynomial~\cite{abramowitz} and
\begin{equation}
  {\cal N}_{n,L} = (-1)^n \left(\frac{\omega}{2}\right)^{\frac{1}{2}(L+\frac{3}{2})} \left(
  \frac{2\, n!}{\Gamma(n+L+\frac{3}{2})}\right)^{1/2}  \label{eq:HO-norm}
\end{equation}
is a normalization coefficient \cite{footnote2}.\par
%
%
All the wavefunctions (\ref{eq:HO-psi}), corresponding to a given value of $L$ and $n=0$, 1, 2,~\ldots, belong to a single positive-discrete series unirrep $D^+_k$ of an su(1,1) Lie algebra. The latter is generated by the operators \cite{wybourne}
\begin{equation}
\begin{split}
  K_0 &= \frac{1}{2\omega} \left(- \frac{d^2}{dr^2} + \frac{L(L+1)}{r^2} + \frac{1}{4} \omega^2 r^2\right)
    = \frac{1}{2\omega} H_{\rm HO}, \\
  K_{\pm} &= \frac{1}{2\omega} \left[\frac{d^2}{dr^2} - \frac{L(L+1)}{r^2} + \frac{1}{4} \omega^2 r^2
    \mp \omega \left(r \frac{d}{dr} + \frac{1}{2}\right)\right],  
\end{split} \label{eq:HO-gen}
\end{equation}
satisfying the commutation relations
\begin{equation}
  [K_0, K_{\pm}] = \pm K_{\pm}, \qquad [K_+, K_-] = - 2K_0,  \label{eq:HO-com}
\end{equation}
and the Hermiticity properties
\begin{equation}
  K_0^{\dagger} = K_0, \qquad K_{\pm}^{\dagger} = K_{\mp},  \label{eq:HO-Hermite}
\end{equation}
while its Casimir operator reads
\begin{equation}
  C_{\rm HO} = - K_+ K_- + K_0 (K_0 - 1).  \label{eq:HO-Casimir}
\end{equation}
The lowest weight characterizing the unirrep is here
\begin{equation}
  k = \tfrac{1}{2} \left(L + \tfrac{3}{2}\right).  \label{eq:HO-k}
\end{equation}
\par
%
%
The wavefunctions $\psi_{n,L}(r)$, $n=0$, 1, 2,~\ldots, are simultaneous eigenfunctions of $C_{\rm HO}$ and $K_0$,
\begin{equation}
\begin{split}
  C_{\rm HO} \psi_{n,L}(r) &= k(k-1) \psi_{n,L}(r) = \tfrac{1}{4} \left(L + \tfrac{3}{2}\right)\left(L - 
    \tfrac{1}{2}\right) \psi_{n,L}(r), \\
  K_0 \psi_{n,L}(r) &= \mu \psi_{n,L}(r) = (k+n) \psi_{n,L}(r) = \frac{1}{2\omega} E_{n,L} \psi_{n,L}(r).
\end{split}  \label{eq:HO-diag}
\end{equation}
Furthermore, $K_+$ and $K_-$ act on them as
\begin{equation}
\begin{split}
  K_+ \psi_{n,L}(r) &= [(\mu-k+1)(\mu+k)]^{1/2} \psi_{n+1,L}(r) \\
  & = \left[(n+1) \left(n+L+\tfrac{3}{2}\right)\right]^{1/2} \psi_{n+1,L}(r), \\
  K_- \psi_{n,L}(r) &= [(\mu-k)(\mu+k-1)]^{1/2} \psi_{n-1,L}(r) = \left[n \left(n+L+\tfrac{1}{2}\right)
    \right]^{1/2} \psi_{n-1,L}(r).
\end{split}  \label{eq:HO-ME}
\end{equation}
\par
%
%
The lowest-energy wavefunction $\psi_{0,L}(r)$ can be obtained as the normalizable solution of the equation $K_- \psi_{0,L}(r) = 0$, {\sl i.e.}, the solution of this equation that vanishes at both points 0 and $\infty$. The remaining wavefunctions $\psi_{n,L}(r)$, $n=1$, 2,~\ldots, can then be built by repeated applications of $K_+$ on $\psi_{0,L}(r)$. Hence, su(1,1) is a spectrum generating algebra for the radial harmonic oscillator.\par
%
%
\subsection{Oscillator-Morse transformation}

Let us now make the changes of variable and of function
\begin{equation}
  r = e^{-x/2}, \qquad \psi_{n,L}(r) = e^{-x/4} \phi_{n,A_0}(x),  \label{eq:HO-M-PCT}
\end{equation}
where $A_0$ is some (yet undetermined) parameter. The coordinate transformation in (\ref{eq:HO-M-PCT}) maps the half-line $0 < r < \infty$ on the full one $- \infty < x < \infty$.\par
%
%
Equation (\ref{eq:HO-SE}) gets modified to the form
\begin{equation}
  H_{{\rm M},n} \phi_{n,A_0}(x) = \epsilon \phi_{n,A_0}(x),  \label{eq:M-SE}
\end{equation}
where
\begin{equation}
  H_{{\rm M},n} = - \frac{d^2}{dx^2} + B^2 e^{-2x} - B (2A_n+1) e^{-x}  \label{eq:M-H}
\end{equation}
and the energy eigenvalue is fixed by
\begin{equation}
  \epsilon = - A_0^2 = - \tfrac{1}{4} \left(L + \tfrac{1}{2}\right)^2.  \label{eq:M-E}
\end{equation}
In (\ref{eq:M-H}), which represents a Morse Hamiltonian \cite{morse, cooper}, one of the parameters
\begin{equation}
  A_n = A_0 + n, \qquad A_0 = \tfrac{1}{2} \left(L + \tfrac{1}{2}\right) = \sqrt{|\epsilon|},  \label{eq:M-A} 
\end{equation}
is $n$-dependent, while the other
\begin{equation}
  B = \tfrac{1}{4} \omega  \label{eq:M-B}
\end{equation}
remains constant.\par
%
%
It is important to stress that in such a map, a \emph{single Hamiltonian} $H_{\rm HO}$, corresponding to a radial harmonic oscillator with a given frequency $\omega$ and a given $L$ value, is transformed into a \emph{hierarchy of Hamiltonians} $H_{{\rm M},n}$, $n=0$, 1, 2,~\ldots, belonging to the Morse family and specified by parameters $(A_n, B)$, with $A_n = A_0 + n$, $n=0$, 1, 2,~\ldots, and constant $A_0$, $B$.\par
%
%
The wavefunctions $\phi_{n,A_0}(x)$ in (\ref{eq:M-SE}) can also be obtained by applying transformation (\ref{eq:HO-M-PCT}) to the harmonic oscillator wavefunctions $\psi_{n,L}(r)$ and are given by
\begin{equation}
  \phi_{n,A_0}(x) = {\cal N}_{n,A_0} \exp\left(- A_0 x - B e^{-x}\right) L_n^{(2A_0)}\left(2B e^{-x}\right),
  \label{eq:M-phi}
\end{equation}
where
\begin{equation}
  {\cal N}_{n,A_0} = (-1)^n (2B)^{A_0 + \frac{1}{2}} \left(\frac{2\, n!}{\Gamma(n+2A_0+1)}\right)^{1/2}.
  \label{eq:M-norm}
\end{equation}
\par
%
%
If, for a moment, we forget the map and focus on a single Hamiltonian of the Morse family with given values of $A_n = \bar{A}$ and $B$, it is clear that it appears in a finite number of Schr\"odinger equations (\ref{eq:M-SE}) since for $A_0$ in (\ref{eq:M-E}), we may choose any of the values $\bar{A} - \bar{n}$ with $\bar{n} = 0$, 1,~\ldots, $\bar{n}_{\rm max}$ ($\bar{A}-1 \le \bar{n}_{\rm max} < \bar{A}$). Hence the resulting energy spectrum $- (\bar{A} - \bar{n})^2$, $\bar{n} = 0$, 1,~\ldots, $\bar{n}_{\rm max}$, coincides with the standard one \cite{morse, cooper}. The same is true for the wavefunctions (\ref{eq:M-phi}) with $A_0 \to \bar{A} - \bar{n}$, except for the fact that the normalization coefficient (\ref{eq:M-norm}) corresponds to a `new' scalar product differing from the standard one. From (\ref{eq:HO-sp}) and (\ref{eq:HO-M-PCT}), we indeed get \cite{footnote3}
\begin{equation}
  \frac{1}{2} \int_{-\infty}^{+\infty} |\phi(x)|^2 e^{-x} dx = 1.  \label{eq:M-sp}
\end{equation}
\par
%
%
Going back to the map (\ref{eq:HO-M-PCT}), we observe that it transforms the harmonic oscillator su(1,1) generators (\ref{eq:HO-gen}) into another set of su(1,1) generators
\begin{equation}
  M_i = e^{x/4} K_i e^{-x/4}, \qquad i = 0, +, -,  \label{eq:HO-M-gen}
\end{equation}
satisfying equations similar to (\ref{eq:HO-com}) and (\ref{eq:HO-Hermite}) (where in the latter the adjoint is defined with respect to the `new' scalar product). Explicitly,
\begin{equation}
\begin{split}
  M_0 &= \frac{1}{2B} e^x \left(- \frac{d^2}{dx^2} + B^2 e^{-2x} - \epsilon\right) = \frac{1}{2B} \left[e^x 
    (H_{{\rm M},n} - \epsilon) + B (2A_n + 1)\right], \\
  M_{\pm} &= \frac{1}{2B} e^x \left(\frac{d^2}{dx^2} + B^2 e^{-2x} + \epsilon\right) \pm \left(\frac{d}{dx}
    - \frac{1}{2}\right),  
\end{split} \label{eq:M-gen}
\end{equation}
with a corresponding Casimir operator
\begin{equation*}
  C_{\rm M} = - M_+ M_- + M_0 (M_0-1).
\end{equation*}
\par
%
%
Comparison between (\ref{eq:HO-k}) and (\ref{eq:M-A}) directly shows that the lowest weight characterizing the unirrep is now
\begin{equation}
  k = A_0 + \tfrac{1}{2} = \sqrt{|\epsilon|} + \tfrac{1}{2}.  \label{eq:M-k}
\end{equation}
Furthermore, the counterparts of Eqs.\ (\ref{eq:HO-diag}) and (\ref{eq:HO-ME}) read
\begin{equation}
\begin{split}
  C_{\rm M} \phi_{n,A_0}(x) &= \left(A_0^2 - \tfrac{1}{4}\right) \phi_{n,A_0}(x) = \left(|\epsilon| - 
    \tfrac{1}{4} \right) \phi_{n,A_0}(x), \\
  M_0 \phi_{n,A_0}(x) &= \left(A_0 + n + \tfrac{1}{2}\right) \phi_{n,A_0}(x) =\left(A_n + \tfrac{1}{2}\right) 
    \phi_{n,A_0}(x),
\end{split}  \label{eq:M-diag}
\end{equation}
and
\begin{equation}
\begin{split}
  M_+ \phi_{n,A_0}(x) &= [(n+1) (n+2A_0+1)]^{1/2} \phi_{n+1,A_0}(x) \\
  & = [(n+1) (2A_n-n+1)]^{1/2} \phi_{n+1,A_0}(x), \\
  M_- \phi_{n,A_0}(x) &= [n (n+2A_0)]^{1/2} \phi_{n-1,A_0}(x) = [n (2A_n-n)]^{1/2} \phi_{n-1,A_0}(x),
\end{split}  \label{eq:M-ME}
\end{equation}
respectively \cite{footnote4}.\par
%
%
It is therefore clear that for the Morse family, the su(1,1) unirrep $D^+_k$, as specified by (\ref{eq:M-k}), is spanned by the eigenfunctions of the family Hamiltonians corresponding to a given energy eigenvalue $\epsilon$. Its lowest-weight state $\phi_{0,A_0}(x)$, annihilated by $M_-$, is associated with the Morse potential having parameters $(A_0, B)$, while the higher-weight states $\phi_{n,A_0}(x)$, $n=1$, 2,~\ldots, obtained from $\phi_{0,A_0}(x)$ by repeated applications of $M_+$, belong to the potentials with parameters $(A_n, B) = (A_0+n, B)$, $n=1$, 2,~\ldots. This behaviour, entirely similar to that of SUSYQM partners for a translational shape-invariant potential \cite{cooper}, is typical of a potential algebra. In usual realizations \cite{alhassid}, such algebras are built by introducing an auxiliary variable, which converts the generators into partial differential operators while preserving their first-order character. In (\ref{eq:M-gen}), on the contrary, the generators remain dependent on a single variable at the expense of becoming second-order differential operators.\par
%
%
\subsection{Morse-Coulomb transformation}

Let us now make further changes of variable and of function
\begin{equation}
  e^{-x} = R, \qquad \phi_{n,A_0}(x) = \frac{1}{\sqrt{R}} \chi_{n,\cl}(R),  \label{eq:M-C-PCT}
\end{equation}
where $R$ runs on the half-line $0 < R < \infty$.\par
%
%
Equation (\ref{eq:M-SE}) leads to
\begin{equation}
  H_{{\rm C},n} \chi_{n,\cl}(R) = \ce \chi_{n,\cl}(R)  \label{eq:C-SE}
\end{equation}
with
\begin{equation*}
  H_{{\rm C},n} = - \frac{d^2}{dR^2} + \frac{\cl(\cl+1)}{R^2} - \frac{2Z_n}{R}
\end{equation*}
and
\begin{equation}
  \cl(\cl+1) = - \epsilon - \tfrac{1}{4} = A_0^2 - \tfrac{1}{4}, \quad Z_n = B \left(A_n + \tfrac{1}{2}\right)   
    = B \left(A_0 + n + \tfrac{1}{2}\right), \quad \ce = - B^2.  \label{eq:C-para} 
\end{equation}
On defining $\cl = \ell + \frac{D-3}{2}$, we see that Eq.\ (\ref{eq:C-SE}) is the radial Schr\"odinger equation for the $D$-dimensional Coulomb problem ($D \ge 2$), where $\ell$ denotes the angular momentum quantum number and the radial wavefunction is given by $R^{-(D-1)/2} \chi_{n,\cl}(R)$ \cite{wybourne, cooper}.\par
%
%
As in the Morse case, we get a \emph{hierarchy of Hamiltonians} $H_{{\rm C},n}$, $n=0$, 1, 2,~\ldots, this time belonging to the Coulomb family and specified by increasing atomic numbers
\begin{equation}
  Z_n = Z_0 \frac{n+\cl+1}{\cl+1}, \qquad n=0, 1, 2, \ldots,  \label{eq:C-Z}
\end{equation}
but a constant $\cl$ value,
\begin{equation}
  \cl = A_0 - \tfrac{1}{2}.  \label{eq:C-L}
\end{equation}
From (\ref{eq:C-para}), it is clear that the energy eigenvalue 
\begin{equation}
  \ce = - \frac{Z_n^2}{(n+\cl+1)^2} = - \frac{Z_0^2}{(\cl+1)^2}  \label{eq:C-E} 
\end{equation}
remains constant again.\par
%
%
{}For the wavefunctions in (\ref{eq:C-SE}), Eqs.\ (\ref{eq:M-phi}), (\ref{eq:M-norm}) and (\ref{eq:M-C-PCT}) yield
\begin{equation}
  \chi_{n,\cl}(R) = {\cal N}_{n,\cl} R^{\cl+1} \exp\left(- \frac{Z_n R}{n+\cl+1}\right) L_n^{(2\cl+1)}
    \left(\frac{2Z_n R}{n+\cl+1}\right) \label{eq:C-chi}
\end{equation}
with
\begin{equation}
  {\cal N}_{n,\cl} = (-1)^n \left(\frac{2Z_n}{n+\cl+1}\right)^{\cl+1} \left(\frac{2\, n!}{\Gamma(n+2\cl+2)}
  \right)^{1/2}.  \label{eq:C-norm} 
\end{equation}
\par
%
%
If we focus on a single Hamiltonian of the Coulomb family, corresponding to given values of $Z_n = \bar{Z}$ and $\cl$, we may associate with it an infinite number of Schr\"odinger equations (\ref{eq:C-SE}), since for $Z_0$ in (\ref{eq:C-E}) we may choose any of the values $\bar{Z}(\cl+1)/(\bar{n}+\cl+1)$, $\bar{n}=0$, 1, 2,~\ldots. This yields the well-known Coulomb energy spectrum $- \bar{Z}^2/(\bar{n}+\cl+1)^2$, $\bar{n}=0$, 1, 2,~\ldots \cite{wybourne, cooper}. The wavefunctions, as given in (\ref{eq:C-chi}) and (\ref{eq:C-norm}), correspond to a `new' scalar product again, because  Eqs.\ (\ref{eq:M-sp}) and (\ref{eq:M-C-PCT}) lead to \cite{footnote5}
\begin{equation*}
  \frac{1}{2} \int_0^{\infty} \frac{|\chi(R)|^2}{R} dR = 1.
\end{equation*}
\par
%
%
Going back to the hierarchy of Coulomb Hamiltonians, we can use the map (\ref{eq:M-C-PCT}) to get a third realization of su(1,1) in terms of
\begin{equation}
  N_i = \sqrt{R}\, M_i \frac{1}{\sqrt{R}}, \qquad i=0, +, -,  \label{eq:M-C-gen}
\end{equation}
or
\begin{equation*}
\begin{split}
  N_0 &= \frac{1}{2\sqrt{|\ce|}} R \left(- \frac{d^2}{dR^2} + \frac{\cl(\cl+1)}{R^2} - \ce\right) = 
    \frac{1}{2\sqrt{|\ce|}} [R (H_{{\rm C},n} - \ce) + 2Z_n], \\
  N_{\pm} &= \frac{1}{2\sqrt{|\ce|}} R \left(\frac{d^2}{dR^2} - \frac{\cl(\cl+1)}{R^2} - \ce\right) \mp R
    \frac{d}{dR},  
\end{split}
\end{equation*}
with a corresponding Casimir operator
\begin{equation*}
  C_{\rm C} = - N_+ N_- + N_0 (N_0-1).
\end{equation*}
\par
%
%
The su(1,1) unirrep is now characterized by
\begin{equation*}
  k = \cl + 1
\end{equation*}
and is spanned by the eigenfunctions of the Coulomb family Hamiltonians corresponding to a given energy eigenvalue $\ce$. The action of the various operators on these eigenfunctions is given by
\begin{equation*}
\begin{split}
  C_{\rm C} \chi_{n,\cl}(R) &= \cl(\cl+1) \chi_{n,\cl}(R), \\
  N_0 \chi_{n,\cl}(R) &= (n+\cl+1) \chi_{n,\cl}(R),
\end{split}  
\end{equation*}
and
\begin{equation}
\begin{split}
  N_+ \chi_{n,\cl}(R) &= [(n+1) (n+2\cl+2)]^{1/2} \chi_{n+1,\cl}(R), \\
  N_- \chi_{n,\cl}(R) &= [n (n+2\cl+1)]^{1/2} \chi_{n-1,\cl}(R) .
\end{split}  \label{eq:C-ME}
\end{equation}
\par
%
%
The lowest-weight state $\chi_{0,\cl}(R)$, annihilated by $N_-$, pertains to the family member with atomic number $Z_0$. Repeated applications of $N_+$ on $\chi_{0,\cl}(R)$ yield the higher-weight states $\chi_{n,\cl}(R)$, $n=1$, 2,~\ldots, belonging to the Coulomb potentials with atomic numbers $Z_n$, $n=1$, 2,~\ldots, given in (\ref{eq:C-Z}). This leads to a picture entirely similar to that previously observed for the Morse hierarchy. We conclude that in the present case too, su(1,1) behaves as a potential algebra.\par
%
%
Before proceeding to PDM Schr\"odinger equations in Sec.\ III, it is worth mentioning that the transformations (\ref{eq:HO-M-PCT}) and (\ref{eq:M-C-PCT}) can be combined to get a direct link between the $d$-dimensional radial harmonic oscillator and the $D$-dimensional radial Coulomb problems. Since this map has been studied in detail in Ref.\ \cite{kostelecky}, it will not be commented here any further.\par
%
%
\section{MAPPINGS AND ALGEBRAS FOR POSITION-DEPENDENT MASS}
\setcounter{equation}{0}

\subsection{Oscillator case}

In Refs.\ \cite{bagchi05} and \cite{cq07}, we considered counterparts of Eqs.\ (\ref{eq:HO-SE}) and (\ref{eq:HO-H}), describing a $d$-dimensional radial harmonic oscillator ($d \ge 2$) in a PDM background. They were obtained by replacing the radial momentum $- {\rm i} d/dr$ in (\ref{eq:HO-H}) by some deformed momentum $\pi_{\rm HO} = - {\rm i} \sqrt{f_{\rm HO}(\alpha; r)}\, (d/dr) \sqrt{f_{\rm HO}(\alpha; r)}$, where $f_{\rm HO}(\alpha; r) = 1 + \alpha r^2$ and $\alpha$ is a positive real constant, yielding
\begin{equation}
  \ha_{\rm HO} \psia_{n,L}(r) = \ea_{n,L} \psia_{n,L}(r)  \label{eq:HO-SE-PDM}
\end{equation}
and
\begin{equation*}
  \ha_{\rm HO} = \pi_{\rm HO}^2 + \frac{L(L+1)}{r^2} + \frac{1}{4} \omega^2 r^2.
\end{equation*}
Equation (\ref{eq:HO-SE-PDM}) can indeed be rewritten in the standard form of a PDM Schr\"odinger equation (see Refs.\ \cite{bagchi04} and \cite{cq06})
\begin{equation}
  \left(- \frac{d}{dr} \frac{1}{M_{\rm HO}(\alpha;r)} \frac{d}{dr} + \veff_{\rm HO}(\alpha;r)\right) 
  \psia_{n,L}(r) = \ea_{n,L} \psia_{n,L}(r)  \label{eq:HO-SE-PDM-bis} 
\end{equation}
with a PDM and an effective potential given by
\begin{equation}
  M_{\rm HO}(\alpha; r) = \frac{1}{f_{\rm HO}^2(\alpha; r)} = \frac{1}{(1 + \alpha r^2)^2}  \label{eq:HO-M}
\end{equation}
and
\begin{equation}
  \veff_{\rm HO}(\alpha;r) = \frac{L(L+1)}{r^2} + \frac{1}{4} (\omega^2 - 8 \alpha^2) r^2 - \alpha,
  \label{eq:HO-Veff}
\end{equation}
respectively. Observe that the constant-mass limit corresponds to $\alpha \to 0$, in which case Eq.\ (\ref{eq:HO-SE-PDM-bis}) gives back Eq.\ (\ref{eq:HO-SE}).\par
%
%
As its constant-mass counterpart, such a PDM Schr\"odinger equation has an infinite number of bound-state solutions
\begin{equation}
  \psia_{n,L}(r) = \cna_{n,L} r^{L+1} f_{\rm HO}^{-[\lambda_{\rm HO} + (L+2) \alpha]/(2\alpha)} 
  P_n^{\left(\frac{\lambda_{HO}}{\alpha} - \frac{1}{2}, L + \frac{1}{2}\right)}(t_{\rm HO}), \qquad n=0, 1,
  2, \ldots,  \label{eq:HO-psi-PDM}
\end{equation}
corresponding to the energy eigenvalues
\begin{equation}
\begin{split}
  \ea_{n,L} &= \alpha \left(4n^2 + 4n(L+1) + L + 1 + (4n + 2L + 3) \frac{\lambda_{\rm HO}}{\alpha}\right),
        \qquad n=0, 1, 2, \ldots, \\
  \lambda_{\rm HO} &= \tfrac{1}{2}(\alpha + \Delta_{\rm HO}), \qquad \Delta_{\rm HO} = \sqrt{\omega^2 +
        \alpha^2}. 
\end{split}  \label{eq:HO-E-PDM}
\end{equation}
The energy spectrum is therefore quadratic instead of linear. The wavefunctions in (\ref{eq:HO-psi-PDM}) are expressed in terms of Jacobi polynomials \cite{abramowitz} in the variable
\begin{equation*}
  t_{\rm HO} = 1 - \frac{2}{f_{\rm HO}} = \frac{-1 + \alpha r^2}{1 + \alpha r^2}  \label{eq:t}
\end{equation*}
and of a normalization coefficient
\begin{equation*}
  \cna_{n,L} = \left(\frac{2 \alpha^{L+\frac{3}{2}} n!\, \left(\frac{\lambda_{\rm HO}}{\alpha}+2n+L+1\right)
  \Gamma\left(\frac{\lambda_{\rm HO}}{\alpha}+n+L+1\right)}{\Gamma\left(\frac{\lambda_{\rm HO}}{\alpha}
  +n+\frac{1}{2}\right) \Gamma\left(n+L+\frac{3}{2}\right)}\right)^{1/2}.  
\end{equation*}
\par
%
%
In Ref.\ \cite{cq07}, it was also shown that for this problem one can construct a deformed su(1,1) algebra going over for $\alpha \to 0$ to the standard one, defined in (\ref{eq:HO-gen}). Its generators $\ka_0$, $\ka_+$ and $\ka_-$ can be written as
\begin{equation*}
\begin{split}
  \ka_0 &= \frac{1}{4\lambda_{\rm HO}} \ha_{\rm HO}, \\
  \ka_{\pm} &= \pm \frac{1}{16\lambda_{\rm HO}} \Aa_{{\rm HO},\pm} (\delta_{\rm HO} \pm 1) \sqrt{\frac
       {\delta_{\rm HO} \pm 2}{\delta_{\rm HO}}} = \pm \frac{1}{16\lambda_{\rm HO}} (\delta_{\rm HO} \mp 
       1) \sqrt{\frac{\delta_{\rm HO}}{\delta_{\rm HO} \mp 2}} \Aa_{{\rm HO},\pm},
\end{split}
\end{equation*}
with
\begin{equation*}
\begin{split}
  \Aa_{{\rm HO},\pm} &= - 4 {\rm i} \alpha \left(2 \frac{r}{f_{\rm HO}} \pi_{\rm HO} + {\rm i} t_{\rm HO}\right) 
      - 4\alpha t_{\rm HO} (1 \mp \delta_{\rm HO}) + 4\alpha \frac{\left(\frac{\lambda_{\rm HO}}{\alpha}-L-1
      \right) \left(\frac{\lambda_{\rm HO}}{\alpha} +L\right)}{1 \pm \delta_{\rm HO}}, \\
  \delta_{\rm HO} &= \sqrt{\frac{4\lambda_{\rm HO}}{\alpha} \ka_0 + \frac{\lambda_{\rm HO}}{\alpha}
      \left(\frac{\lambda_{\rm HO}}{\alpha}-1\right) + L(L+1)},
\end{split}
\end{equation*}
and satisfy the commutation relations
\begin{equation}
\begin{split}
  \bigl[\ka_0, \ka_{\pm}\bigr] &= \pm \frac{\alpha}{\lambda_{\rm HO}} \ka_{\pm} (\delta_{\rm HO} \pm 1) 
        = \pm \frac{\alpha}{\lambda_{\rm HO}} (\delta_{\rm HO} \mp 1) \ka_{\pm}, \\
  \bigl[\ka_+, \ka_-\bigr] &= - \frac{\alpha\delta_{\rm HO}}{\lambda_{\rm HO}} \left(2 \ka_0 + \frac{\alpha}{4
        \lambda_{\rm HO}}\right),   
\end{split}  \label{eq:HO-com-PDM}
\end{equation}
together with Hermiticity properties similar to (\ref{eq:HO-Hermite}). The corresponding Casimir operator $\ca_{\rm HO}$, analogous to (\ref{eq:HO-Casimir}), reads
\begin{equation}
  \ca_{\rm HO} = - \ka_+ \ka_- + K^{(\alpha)2}_0 - \frac{\alpha}{\lambda_{\rm HO}} \left(\delta_{\rm HO} - 
  \frac{5}{4}\right) \ka_0 - \frac{\alpha^2}{8\lambda_{\rm HO}^2} \delta_{\rm HO}.  \label{eq:HO-Casimir-PDM} 
\end{equation}
\par
%
%
Equations (\ref{eq:HO-diag}) and (\ref{eq:HO-ME}) are now replaced by 
\begin{equation}
\begin{split}
  \ca_{\rm HO} \psia_{n,L} &= \left[\frac{1}{4} \left(1 - \frac{\alpha}{\lambda_{\rm HO}}\right) \left(L + 
       \frac{3}{2}\right) \left(L - \frac{1}{2}\right) - \frac{3\alpha^2}{16\lambda_{\rm HO}^2} L(L+1)\right] 
       \psia_{n,L}, \\
  \ka_0 \psia_{n,L} &= \frac{1}{4\lambda_{\rm HO}} \ea_{n,L} \psia_{n,L},
\end{split}  \label{eq:HO-diag-PDM}
\end{equation}
and
\begin{equation}
\begin{split}
  \ka_+ \psia_{n,L} &= \frac{\alpha}{\lambda_{\rm HO}} \left[(n+1) \left(n+L+\frac{3}{2}\right) \left(n+\frac
     {\lambda_{\rm HO}}{\alpha}+L+1\right) \left(n+\frac{\lambda_{\rm HO}}{\alpha}+\frac{1}{2}\right)\right]
     ^{1/2} \\
  & \quad \times \psia_{n+1,L}, \\
  \ka_- \psia_{n,L} &= \frac{\alpha}{\lambda_{\rm HO}} \left[n \left(n+L+\frac{1}{2}\right) \left(n+\frac
     {\lambda_{\rm HO}}{\alpha}+L\right) \left(n+\frac{\lambda_{\rm HO}}{\alpha}-\frac{1}{2}\right)\right]
     ^{1/2} \psia_{n-1,L}, 
\end{split}  \label{eq:HO-ME-PDM}
\end{equation}
respectively.\par
%
%
{}From (\ref{eq:HO-diag-PDM}) and (\ref{eq:HO-ME-PDM}), it is clear that all the wavefunctions (\ref{eq:HO-psi-PDM}), corresponding to a given $L$ value and $n=0$, 1, 2,~\ldots, belong to a single positive-discrete series unirrep of the deformed su(1,1) algebra generated by $\ka_0$, $\ka_+$ and $\ka_-$. As shown explicitly in Ref.\ \cite{cq07}, the lowest-weight wavefunction $\psia_{0,L}(r)$ is the normalizable solution of the equation $\ka_- \psia_{0,L}(r) = 0$, while the remaining wavefunctions can be built from it by repeated applications of $\ka_+$. This led us to the conclusion that such a deformed su(1,1) algebra provides a spectrum generating algebra for the PDM $d$-dimensional radial harmonic oscillator problem specified by Eqs.\ (\ref{eq:HO-SE-PDM-bis}), (\ref{eq:HO-M}) and (\ref{eq:HO-Veff}).\par
%
%
We are now in a position to combine the outcomes of this subsection with the PCT approach of Sec.\ II to derive some new results.\par
%
%
\subsection{Oscillator-Morse transformation}

Let us apply the transformation (\ref{eq:HO-M-PCT}) to the PDM $d$-dimensional radial harmonic oscillator Schr\"odinger equation (\ref{eq:HO-SE-PDM}). Since the deforming function $f_{\rm HO}(\alpha;r)$ becomes $f_{\rm M}(\alpha;x) = 1 + \alpha e^{-x}$ and the deformed momentum $\pi_{\rm HO}$ is changed into $- 2 e^{x/2} \pi_{\rm M}$, with $\pi_{\rm M}$ defined by $\pi_{\rm M} = - {\rm i} \sqrt{f_{\rm M}(\alpha; x)}\, (d/dx) \sqrt{f_{\rm M}(\alpha; x)}$, we arrive, after some lengthy but straightforward calculations, at an equation
\begin{equation}
  \ha_{{\rm M},n} \phia_{n,A_0}(x) = \epsilona \phia_{n,A_0}(x)  \label{eq:M-SE-PDM}
\end{equation}
for a PDM Morse Hamiltonian
\begin{equation}
  \ha_{{\rm M},n} = \pi_{\rm M}^2 + B^2 e^{-2x} - B (2A_n+1) e^{-x},  \label{eq:M-H-PDM}
\end{equation}
which can be written in a form similar to (\ref{eq:HO-SE-PDM-bis}) with
\begin{equation}
  M_{\rm M}(\alpha; x) = \frac{1}{f_{\rm M}^2(\alpha; x)} = \frac{1}{(1 + \alpha e^{-x})^2}  \label{eq:M-M}
\end{equation}
and
\begin{equation*}
  \veff_{\rm M}(\alpha;x) = \left(B^2 - \frac{3}{4} \alpha^2\right) e^{-2x} - \left[B (2A_n+1) + \frac{\alpha}{2}
  \right] e^{-x}. 
\end{equation*}
\par
%
%
The relations between the parameters $\omega$, $L$ and the energy eigenvalue $\ea_{n,L}$ of the harmonic oscillator and the corresponding quantities $A_n$, $B$, and $\epsilona$ for the Morse potential, however, turn out to be more complicated than in the constant-mass case and are given by
\begin{equation*}
  B = \frac{1}{4} \sqrt{\omega^2 - 3\alpha^2}, \qquad 2A_n+1 = \frac{\ea_{n,L} + \frac{\alpha}{2}}{\sqrt{
  \omega^2 - 3\alpha^2}}, \qquad \epsilona = - \frac{1}{4} \left(L + \frac{1}{2}\right)^2.
\end{equation*}
While the Morse energy eigenvalue $\epsilona$ and potential parameter $B$ remain constant and rather similar to their constant-mass counterparts (\ref{eq:M-E}) and (\ref{eq:M-B}), the other potential parameter
\begin{align}
 A_n &= \frac{\alpha n + |\lambda_{\rm M}|}{|\lambda_{\rm M}|} A_0 + n \frac{2B^2 + \alpha B + \alpha 
      |\lambda_{\rm M}| (n+1)}{2B |\lambda_{\rm M}|},  \label{eq:M-A-PDM} \\
 A_0 &= \frac{1}{2} \left(\frac{(2L+3)|\lambda_{\rm M}|}{2B} - 1\right),  \label{eq:M-A0-PDM}
\end{align}
gets a quadratic $n$-dependence instead of a linear one, due to Eq.\ (\ref{eq:HO-E-PDM}). In (\ref{eq:M-A-PDM}), we have defined
\begin{equation*}
  \lambda_{\rm M} = - \frac{1}{2} (\alpha + \Delta_{\rm M}) = - \frac{1}{2} \left(\lambda_{\rm HO} + \frac{
  \alpha}{2}\right), \qquad \Delta_{\rm M} = \sqrt{4B^2 + \alpha^2} = \frac{1}{2} \Delta_{\rm HO}. 
\end{equation*}
As in the constant-mass case, a \emph{single Hamiltonian} $\ha_{\rm HO}$ with given values of $\omega$ and $L$ is mapped onto a \emph{hierarchy of Hamiltonians} $\ha_{{\rm M},n}$ with constant $B$ but varying $A_n$.\par
%
%
The Morse wavefunctions $\phia_{n,A_0}(x)$ can also be derived from (\ref{eq:HO-M-PCT}) and (\ref{eq:HO-psi-PDM}) and read
\begin{equation}
  \phia_{n,A_0}(x) = \cna_{n,A_0} e^{- \sqrt{|\epsilona|}\, x} f_{\rm M}^{- \left[|\lambda_{\rm M}| + \left(\sqrt{
  |\epsilona|} + \frac{1}{2}\right) \alpha\right]/\alpha} P_n^{\left(2 \frac{|\lambda_{\rm M}|}{\alpha} - 1, 2
  \sqrt{|\epsilona|}\right)} (t_{\rm M}),  \label{eq:M-phi-PDM}
\end{equation}
where
\begin{equation*}
  t_{\rm M} = 1 - \frac{2}{f_{\rm M}} = \frac{- 1 + \alpha e^{-x}}{1 + \alpha e^{-x}}
\end{equation*}
and
\begin{equation*}
  \cna_{n,A_0} = \left(\frac{2 \alpha^{2\sqrt{|\epsilona|} + 1} n!\, \left(2 \frac{|\lambda_{\rm M}|}{\alpha} + 2n
  + 2\sqrt{|\epsilona|}\right) \Gamma\left(2 \frac{|\lambda_{\rm M}|}{\alpha} + n + 2\sqrt{|\epsilona|}\right)}
  {\Gamma\left(2 \frac{|\lambda_{\rm M}|}{\alpha} + n\right) \Gamma\left(n + 2\sqrt{|\epsilona|} + 1\right)}
  \right)^{1/2}.
\end{equation*}
\par
%
%
We observe that a given Hamiltonian (\ref{eq:M-H-PDM}) of the Morse family with constant values of $A_n = \bar{A}$ and $B$ may appear in several equations (\ref{eq:M-SE-PDM}), because for $A_0$ in
\begin{equation*}
  \epsilona = - \frac{1}{4} \left(\frac{(2A_0+1)B - |\lambda_{\rm M}|}{|\lambda_{\rm M}|}\right)^2,
\end{equation*}
we may choose distinct values satisfying  Eq.\ (\ref{eq:M-A-PDM}) with $A_n$ and $n$ replaced by $\bar{A}$ and $\bar{n}$, respectively. As a result, we get an energy spectrum $- [(2\bar{A}+1)B - \alpha \bar{n}^2 - (2\bar{n}+1) |\lambda_{\rm M}|]^2 [2 (\alpha \bar{n} + |\lambda_{\rm M}|)]^{-2}$, where $\bar{n}=0$, 1,~\ldots, $\bar{n}_{\rm max}$ and $\bar{A}-1 \le \bar{n}_{\rm max} < \bar{A}$, provided $\alpha$ remains small enough (see Ref.\ \cite{bagchi05} for more details) \cite{footnote6}.\par
%
%
Going back to the hierarchy of PDM Morse Hamiltonians (\ref{eq:M-H-PDM}), we can construct for it a deformed su(1,1) algebra by a transformation similar to Eq.\ (\ref{eq:HO-M-gen}). The resulting generators can be written as 
\begin{equation*}
\begin{split}
  \ma_0 &= \frac{2}{4|\lambda_{\rm M}| - \alpha} e^x \left(\pi_{\rm M}^2 + B^2 e^{-2x} - \frac{\alpha}{8} 
        e^{-x} - \epsilona\right) \\
  &= \frac{2}{4|\lambda_{\rm M}| - \alpha} \left[e^x \left(\ha_{{\rm M},n} - \epsilona\right) + B(2A_n+1)
        - \frac{\alpha}{8}\right], \\
  \ma_{\pm} &= \pm \frac{1}{8(4|\lambda_{\rm M}| - \alpha)} \Aa_{{\rm M},\pm} (\delta_{\rm M} \pm 1) 
        \sqrt{\frac{\delta_{\rm M} \pm 2}{\delta_{\rm M}}} = \pm \frac{1}{8(4\lambda_{\rm M}| - \alpha)} 
        (\delta_{\rm M} \mp 1) \sqrt{\frac{\delta_{\rm M}}{\delta_{\rm M} \mp 2}} \Aa_{{\rm M},\pm},\end{split}
\end{equation*}
with
\begin{equation*}
\begin{split}
  \Aa_{{\rm M},\pm} &= \frac{8 {\rm i} \alpha}{f_{\rm M}} (2 \pi_{\rm M} + {\rm i}) - 4\alpha t_{\rm M} 
      (1 \mp \delta_{\rm M}) + 4\alpha \frac{\left(2 \frac{|\lambda_{\rm M}|}{\alpha} - 2 \sqrt{|\epsilona|} - 1
      \right) \left(2 \frac{|\lambda_{\rm M}|}{\alpha} + 2 \sqrt{|\epsilona|} - 1\right)}{1 \pm \delta_{\rm M}}, \\
  \delta_{\rm M} &= \sqrt{\frac{2(4|\lambda_{\rm M}| - \alpha)}{\alpha} \ma_0 + 4 \frac{|\lambda_{\rm M}|}
      {\alpha} \left(\frac{|\lambda_{\rm M}|}{\alpha} - 1\right) + 4 |\epsilona| + \frac{1}{2}}.
\end{split}
\end{equation*}
The deformed commutation relations and the Casimir operator take a form similar to Eqs.\ (\ref{eq:HO-com-PDM}) and (\ref{eq:HO-Casimir-PDM}) with $\alpha/\lambda_{\rm HO}$ and $\delta_{\rm HO}$ replaced by $2\alpha/(4|\lambda_{\rm M}|-\alpha)$ and $\delta_{\rm M}$, respectively.\par
%
%
The introduction of the PDM (\ref{eq:M-M}) therefore changes Eqs.\ (\ref{eq:M-diag}) and (\ref{eq:M-ME}) into
\begin{equation*}
\begin{split}
  \ca_{\rm M} \phia_{n,A_0} &= \left[\left(1 - \frac{2\alpha}{4|\lambda_{\rm M}|-\alpha}\right) \left(|\epsilona| 
      - \frac{1}{4}\right) - \frac{3\alpha^2}{(4|\lambda_{\rm M}|-\alpha)^2} \left(|\epsilona| - \frac{1}{16}\right) 
      \right] \phia_{n,A_0}, \\
  \ma_0 \phia_{n,A_0} &= \frac{2}{4|\lambda_{\rm M}|-\alpha} \left[(2A_0+1)B + 2|\lambda_{\rm M}|n +
     \alpha \left(n(n-1) + n \frac{(2A_0+1)B}{|\lambda_{\rm M}|} - \frac{1}{8}\right)\right] \phia_{n,A_0},  
\end{split}
\end{equation*}
and
\begin{equation*}
\begin{split}
  \ma_+ \phia_{n,A_0} &= \frac{2}{4|\lambda_{\rm M}|-\alpha} \left[(n+1) \left(n + \frac{(2A_0+1)B}
     {|\lambda_{\rm M}|}\right) \right]^{1/2} \\
  & \quad \times \left\{\left[2|\lambda_{\rm M}| + \alpha \left(n + \frac{(2A_0+1)B}{|\lambda_{\rm M}|}
     - 1\right)\right](2|\lambda_{\rm M}| + \alpha n)\right\}^{1/2} \phia_{n+1,A_0}, \\   
  \ma_- \phia_{n,A_0} &= \frac{2}{4|\lambda_{\rm M}|-\alpha} \left[n \left(n + \frac{(2A_0+1)B}
     {|\lambda_{\rm M}|} - 1\right) \right]^{1/2} \\
  & \quad \times \left\{\left[2|\lambda_{\rm M}| + \alpha \left(n + \frac{(2A_0+1)B}
     {|\lambda_{\rm M}|} - 2\right)\right] [2|\lambda_{\rm M}| + \alpha (n-1)]\right\}^{1/2} \phia_{n-1,A_0}.
\end{split}
\end{equation*}
These new relations, though exhibiting a more complicated $n$ dependence than the old ones, have exactly the same interpretation so that we have an algebra belonging to the potential algebra class again.\par
%
%
\subsection{Morse-Coulomb transformation}

{}Finally, let us apply the transformation (\ref{eq:M-C-PCT}) to the PDM Morse Schr\"odinger equation (\ref{eq:M-SE-PDM}). This time, $f_{\rm M}(\alpha;x)$ leads to $f_{\rm C}(\alpha;R) = 1 + \alpha R$, while $\pi_{\rm M}$ is mapped onto $- R \pi_{\rm C}$, with $\pi_{\rm C} = - {\rm i} \sqrt{f_{\rm C}(\alpha; R)}\, (d/dR) \sqrt{f_{\rm C}(\alpha; R)}$. As a result, we get an equation
\begin{equation}
  \ha_{{\rm C},n} \chia_{n,\cl}(R) = \cea \chia_{n,\cl}(R)  \label{eq:C-SE-PDM} 
\end{equation}
for a PDM Coulomb Hamiltonian
\begin{equation*}
  \ha_{{\rm C},n} = \pi_{\rm C}^2 + \frac{\cl(\cl+1)}{R^2} - \frac{2Z_n}{R}.
\end{equation*}
Equation (\ref{eq:C-SE-PDM}) can indeed be rewritten in a form similar to (\ref{eq:HO-SE-PDM-bis}) with a PDM
\begin{equation}
  M_{\rm C}(\alpha;R) = \frac{1}{f_{\rm C}^2(\alpha;R)} = \frac{1}{(1 + \alpha R)^2}  \label{eq:C-M}
\end{equation}
and an effective potential
\begin{equation*}
  \veff_{{\rm C},n}(\alpha;R) = \frac{\cl(\cl+1)}{R^2} - \frac{2Z_n}{R} - \frac{\alpha^2}{4}. 
\end{equation*}
\par
%
%
The relations (\ref{eq:C-para}), (\ref{eq:C-Z}) and (\ref{eq:C-L}) are now replaced by
\begin{equation*}
  \cl(\cl+1) = - \epsilona - \frac{1}{4}, \qquad Z_n = B \left(A_n + \frac{1}{2}\right), \qquad \cea = - B^2 - 
  \frac{\alpha^2}{4},
\end{equation*}
\begin{equation}
  Z_n = Z_0 \frac{n+\cl+1}{\cl+1} + \frac{\alpha}{2} n (n+2\cl+1),  \label{eq:C-Z-PDM}
\end{equation}
and
\begin{equation}
  \cl = \frac{(2A_0+1)B - 2|\lambda_{\rm M}|}{2|\lambda_{\rm M}|},  \label{eq:C-L-PDM}
\end{equation}
respectively, while the wavefunctions $\chia_{n,\cl}(R)$ in (\ref{eq:C-SE-PDM}) can be expressed as
\begin{equation}
  \chia_{n,\cl}(R) = \cna_{n,\cl} R^{\cl+1} f_{\rm C}^{-\left[\sqrt{|\cea|} + \left(\cl + \frac{3}{2}\right)\alpha
  \right]/\alpha} P_n^{\left(\frac{2}{\alpha} \sqrt{|\cea|}, 2\cl+1\right)}(t_{\rm C}),  \label{eq:C-chi-PDM}
\end{equation}
with
\begin{equation*}
  t_{\rm C} = 1 - \frac{2}{f_{\rm C}} = \frac{-1 + \alpha R}{1 + \alpha R}
\end{equation*}
and 
\begin{equation*}
  \cna_{n,\cl} = \left(\frac{2 \alpha^{2\cl+2} n! \left(\frac{2}{\alpha}\sqrt{|\cea|} + 2n + 2\cl + 2\right)
  \Gamma\left(\frac{2}{\alpha}\sqrt{|\cea|} + n + 2\cl + 2\right)}{\Gamma\left(\frac{2}{\alpha}\sqrt{|\cea|} + n +
  1\right) \Gamma(n + 2\cl + 2)}\right)^{1/2}.
\end{equation*}
From Eqs.\ (\ref{eq:C-Z-PDM}) and (\ref{eq:C-L-PDM}), we observe that as in the constant-mass case, the 
resulting \emph{hierarchy of} PDM $D$-dimensional Coulomb \emph{Hamiltonians} $H_{{\rm C},n}$, $n=0$, 1, 2,~\ldots, corresponds to a constant $\cl$ value but increasing atomic numbers $Z_n$, $n=0$, 1, 2,~\ldots.\par
%
%
If we focus on a single member of such a family with $Z_n = \bar{Z}$, in contrast with what happens for constant mass, such a Hamiltonian may only appear in a finite number of PDM Schr\"odinger equations (\ref{eq:C-SE-PDM}), associated with a finite number of bound-state energies $- \{2\bar{Z} - \alpha [\bar{n}^2 + (\cl+1)(2\bar{n}+1)]\}^2 [2(\bar{n}+\cl+1)]^{-2}$, $\bar{n}=0$, 1,~\ldots, $\bar{n}_{\rm max}$ (where $\bar{n}_{\rm max}$ has been defined in Ref.\ \cite{bagchi05}). Hence, the introduction of the PDM (\ref{eq:C-M}) has the effect of converting an infinite number into a finite number of bound states. As previously observed \cite{bagchi05}, the suppression (or, conversely, the generation) of bound states is one of the most striking outcomes that a PDM environment may bring about.\par
%
%
Nevertheless, no consequence ensues from this phenomenon as far as the construction of a deformed su(1,1) algebra is concerned since the latter is based upon the (infinite) hierarchy of Hamiltonians $H_{{\rm C},n}$. Proceeding as in (\ref{eq:M-C-gen}), we therefore obtain the generators
\begin{equation*}
\begin{split}
  \na_0 &= \frac{2}{4\sqrt{|\cea|} + \alpha} R \left(\pi_{\rm C}^2 + \frac{\cl(\cl+1)}{R^2} - \frac{\alpha}{8R} -
     \cea\right) \\
  &= \frac{2}{4\sqrt{|\cea|} + \alpha} \left[R \left(\ha_{{\rm C},n} - \cea\right) + 2Z_n - \frac{\alpha}{8}\right], 
     \\
  \na_{\pm} &= \pm \frac{1}{8 \left(4\sqrt{|\cea|} + \alpha\right)} \Aa_{{\rm C},\pm} (\delta_{\rm C} \pm 1)
     \sqrt{\frac{\delta_{\rm C} \pm 2}{\delta_{\rm C}}} \\
  & = \pm \frac{1}{8 \left(4\sqrt{|\cea|} + \alpha\right)} 
     (\delta_{\rm C} \mp 1) \sqrt{\frac{\delta_{\rm C}}{\delta_{\rm C} \mp 2}} \Aa_{{\rm C},\pm}, 
\end{split}
\end{equation*}
where
\begin{equation*}
\begin{split}
  \Aa_{{\rm C},\pm} &= - \frac{8{\rm i}\alpha}{f_{\rm C}} R (2\pi_{\rm C} + {\rm i}\alpha) - 4\alpha t_{\rm C}
      (1 \mp \delta_{\rm C}) \\
  &\quad + 4\alpha \frac{\left(\frac{2}{\alpha} \sqrt{|\cea|} - 2\cl - 1\right) \left(
      \frac{2}{\alpha} \sqrt{|\cea|} + 2\cl + 1\right)}{1 \pm \delta_{\rm C}}, \\
  \delta_{\rm C} &= \sqrt{\frac{2}{\alpha} \left(4\sqrt{|\cea|} + \alpha\right) \na_0 + \frac{4}{\alpha^2} |\cea| 
      + 4\cl(\cl+1) + \frac{1}{2}}.
\end{split}
\end{equation*}
The corresponding deformed commutation relations and Casimir operator have the same structure as Eqs.\ (\ref{eq:HO-com-PDM}) and (\ref{eq:HO-Casimir-PDM}) with $\alpha/\lambda_{\rm HO}$ and $\delta_{\rm HO}$ replaced by $2\alpha/\left(4\sqrt{|\cea|} + \alpha\right)$ and $\delta_{\rm C}$, respectively.\par
%
%
The action of the various operators on the wavefunctions reads
\begin{equation*}
\begin{split}
  \ca_{\rm C} \chia_{n,\cl} &= \Biggl[\Biggl(1 - \frac{2\alpha}{4\sqrt{|\cea|} + \alpha} - \frac{3\alpha^2}{\bigl(4
        \sqrt{|\cea|} + \alpha\bigr)^2}\Biggr) \cl(\cl+1) \\
  & \quad - \frac{9\alpha^2}{16\bigl(4\sqrt{|\cea|} + \alpha\bigr)^2}\Biggr] \chia_{n,\cl}, \\
  \na_0 \chia_{n,\cl} &= \frac{2}{4Z_0 - \alpha(\cl+1)} \left[2Z_0(n+\cl+1) + \alpha(\cl+1)\left(n(n+2\cl+1)
        - \frac{1}{8}\right)\right] \chia_{n,\cl}, 
\end{split}
\end{equation*}
and
\begin{equation*}
\begin{split}
  \na_+ \chia_{n,\cl} &= \frac{2}{4Z_0 - \alpha(\cl+1)} \{(n+1) (n+2\cl+2) [2Z_0 + \alpha(\cl+1)(n+2\cl+1)]\}
       ^{1/2} \\
  & \quad \times [2Z_0 + \alpha(\cl+1)n]^{1/2} \chia_{n+1,\cl}, \\
  \na_- \chia_{n,\cl} &= \frac{2}{4Z_0 - \alpha(\cl+1)} \{n (n+2\cl+1) [2Z_0 + \alpha(\cl+1)(n+2\cl)]\}^{1/2} \\
  & \quad \times [2Z_0 + \alpha(\cl+1)(n-1)]^{1/2} \chia_{n-1,\cl},
\end{split}
\end{equation*}
thereby establishing that we have here an example of potential algebra again.\par
%
%
Before concluding, it is worth commenting on the physical interpretation of the mappings both for constant mass and for PDM. Since $L$ and $\cl$ are related to the angular momentum quantum numbers $l$ and $\ell$, respectively, they assume only integer (resp.\ half-integer) values for odd (resp.\ even) dimension. In view of Eqs.\ (\ref{eq:M-A}), (\ref{eq:C-L}), (\ref{eq:M-A0-PDM}) and (\ref{eq:C-L-PDM}), this implies some restrictions on the Morse parameter 
$A_0$ or on the set of Morse and deforming parameters $(A_0, B, \alpha)$ if one wants to apply the present formalism to some physical problem. On the contrary, the fact that $Z_n$ may not be necessarily integer is no problem since in atomic physics it is customary to use effective atomic numbers.\par
%
%
\section{CONCLUSION}

In this paper, we have reviewed the $d$-dimensional radial harmonic oscillator, the Morse and the $D$-dimensional radial  Coulomb Schr\"odinger equations and provided a unified approach to their bound-state solutions and su(1,1) description by applying the PCT method.\par
%
%
We have demonstrated that in this process the harmonic oscillator su(1,1) spectrum generating algebra is converted into a potential algebra when going to the Morse and the Coulomb problems. In the generators of the latter, the usual presence of an auxiliary variable has been avoided by resorting to second-order differential operators instead of first-order ones.\par
%
%
{}Furthermore, we have shown that this type of analysis can be extended to the cases where the three Schr\"odinger equations contain some well-chosen PDM. This has allowed us to easily generalize a deformed su(1,1) construction, recently achieved for a PDM $d$-dimensional radial harmonic oscillator equation, to the Morse and $D$-dimensional Coulomb potentials. We have established that the PDM presence and the resulting deformation do not change the categorization of the algebra in the spectrum generating or potential algebra class.\par
%
%
{}Finally, it is worth observing that the Morse and Coulomb potentials provide us with the first examples of equivalence between the deformed shape invariance approach proposed in Ref.\ \cite{bagchi05} and a description in terms of a deformed su(1,1) potential algebra, thereby extending the results of Ref.\ \cite{gango} to a PDM background. Whether such a relationship remains true for other potential and PDM pairs is an interesting open problem for future investigation.\par
%
%


\begin{thebibliography}{99}
\bibitem{footnote1} Note, however, that su(1,1) has also been signalled in the context of more general satellite algebras (see Ref.\ \cite{delsol}).
\bibitem{footnote2} The optional phase factor $(-1)^n$ in (\ref{eq:HO-norm}) is not present in Eq.\ (28.5) of Ref.\ \cite{moshinsky96}. We need it here to get positive matrix elements for $K_+$ and $K_-$ (see Eq.\ (\ref{eq:HO-ME})).
\bibitem{footnote3} It is easy to see that the standard wavefunctions can be obtained from (\ref{eq:M-phi}) by multiplying the latter by the $n$-independent factor $[A_0/(2B)]^{1/2}$.
\bibitem{footnote4} Since $\phi_{n,A_0}(x)$ and the standard wavefunctions only differ by an $n$-independent factor, Eq.\ (\ref{eq:M-ME}) remains true for the latter.
\bibitem{footnote5} The wavefunctions corresponding to the standard normalization can be obtained from $\chi_{n,\cl}(R)$ by multiplying the latter by $|\ce|^{1/4} [2(n+\cl+1)]^{-1/2}$. Since this factor depends on $n$, it modifies the matrix elements of the su(1,1) generators, as given in (\ref{eq:C-ME}) (see also Ref.\ \cite{moshinsky72}).
\bibitem{footnote6} As in the constant-mass limit, the wavefunctions (\ref{eq:M-phi-PDM}) and (\ref{eq:C-chi-PDM}) are normalized according to some `new' scalar product differing from the usual one.
\end{thebibliography}
\end{document}